\documentclass[%
 reprint,
amsmath,amssymb,
 aps,
]{revtex4-1}
\usepackage{amsmath}    
\usepackage{amsmath,amssymb}
\usepackage{graphicx}   
\usepackage{verbatim}   
\usepackage{color}      
\usepackage{bm}
\usepackage{subfigure}  
\usepackage{hyperref}   
\usepackage[normalem]{ulem} 
\usepackage{siunitx} 
\usepackage[pagewise]{lineno}

\renewcommand{\figurename}{FIG.}
\renewcommand{\refname}{References}

\begin{document}
\title{Randomness and optimality in enhanced DNA ligation with crowding effects}
\author{Takaharu Y. Shiraki$^1$, Ken-ichiro Kamei$^2$, and Yusuke T. Maeda$^{1}$}
\affiliation{$^1$Kyushu University, Department of Physics, Motooka 744, Fukuoka 812-0395, Japan}
\affiliation{$^2$Kyoto University, Institute for Integrated Cell-Material Sciences (iCeMS), Yoshida-Ushinomiyacho, Sakyo-ku, Kyoto 606-8501, Japan}

\date{\today}

\begin{abstract}
Enzymatic ligation is essential for the synthesis of long DNA. However, the number of ligated products exponentially decays as the DNA synthesis proceeds in a random manner. The controlling of ligation randomness is of importance to suppress exponential decay and demonstrate an efficient synthesis of long DNA. Here, we report the analysis of randomness in sequential DNA ligations, named qPCR-\textit{\underline{b}ased} \underline{ST}atistical \underline{A}nalysis of \underline{R}andomness (qPCR-\textit{b}STAR), by a probability distribution of ligated DNA concentration. We show that the exponential decay is suppressed in a solution of another polymer and DNA ligation is activated at an optimal crowded condition. Theoretical model of kinetic ligation explains that intermolecular attraction due to molecular crowding can be involved in the optimal balance of the ligation speed and the available ligase. Our finding indicates that crowding effects enhance the synthesis of long DNA that retains large genetic information.
\end{abstract}

\maketitle

\section*{Introduction}
Genetic polymers such as DNA or RNA need to be elongated in order to store information in their sequences. To make a long DNA polymer, sequential ligation of DNA fragment catalyzed by a ligase enzyme is an essential step for the repairing DNA break of genome in living cells and the \textit{in vitro} synthesis of the artificial genome\cite{alberts}\cite{phillips}\cite{clyde}. As end-to-end ligation occurs in a test tube is random process, new short fragments are connected from both ends to make longer DNA but the concentration of ligated polymers decays with the number of newly added monomers \cite{Reddy}\cite{landau}. Because ligation randomness restricts the abundance of the polymerized product, it has been long discussed how long genetic polymers are synthesized in nature and technology through random enzymatic ligation.

One scenario to drive an efficient synthesis of long DNA is the local increase of short DNA fragments by using physical transport effects. It has been studied that DNA fragments whose size range from a few tens to a few hundreds of base pairs are accumulated due to molecular transport and thermal convection under temperature gradients\cite{braun1}\cite{braun2}\cite{braun3}\cite{priye}. A local temperature gradient or solute concentration gradient induces the directed motion of DNA as a solute and in turn the trapped DNA shows an exponential increase in the concentration\cite{maeda1}\cite{maeda2}. This DNA trapping is thought to compensate the significant decrease of long DNA polymers. 

The other relevant scenario is the effect from the coexisted polymer, known as molecular crowding\cite{zimmerman}. When DNA coexist with large concentrations of polymers with a smaller gyration radius, attractive interactions occur among DNA due to an excluded volume effect\cite{marenduzzo}\cite{asakura}. This molecular crowding has a wide range of effects from structural changes and catalytic activity of protein enzymes\cite{kilburn}\cite{kim}\cite{kang}\cite{akabayov} to subdiffusive motion\cite{weiss}\cite{sokolov}. The interplay between molecular crowding and the DNA ligation is of importance, but the lack of experimental methods to measure and control the randomness of DNA ligation is major bottleneck to reveal the physical mechanism for long DNA synthesis in crowded condition.

In this study, we investigate the effect of molecular crowding on the long DNA synthesis by using end-to-end ligation of short DNA fragments with coexisting polymer (polyethylene glycol, PEG). To reveal the randomness in ligation, the concentration of ligated DNA products is measured by a quantitative PCR (qPCR)-based method. This method quantifies the concentration distribution of ligated DNA products with the number of ligated joints per molecule. We find that the ligated DNA fragments show the gradual exponential decay in the presence of crowding agents, as the number of ligated joints increases and the production of long DNA is much enhanced with smaller decay rate. We further test the optimal condition of molecular crowding by changing the PEG concentration and find that the exponential decay becomes larger and the ligation efficiency is decreased due to the reduction of freely available ligase as PEG concentration is larger than 10\%. The control of such ligation randomness may be useful not only for the synthesis of artificial genomes \textit{in vitro} but also the understanding of enzymatic reactions in crowded conditions such as intracellular space.

\section*{Results}

\subsection*{The analysis of randomness in enzymatic ligation}
We examine the enzymatic ligation of short DNA fragments under macromolecular crowding. To analyze how the long DNA strands are efficiently synthesized in crowding condition, an analytical method that can analyze a ligated DNA product on its length and concentration is needed. This technical requirement motivates us to develop new method for probing the probability distribution of DNA concentration with respect to size, named as the qPCR-\textit{based} Statistical Analysis of Randomness (qPCR-\textit{b}STAR).

\begin{figure}[tbp]
\centering
\vspace*{10.5em} 
\vspace*{-10em} 
\hspace*{-1em} 
            \includegraphics[scale=0.13]{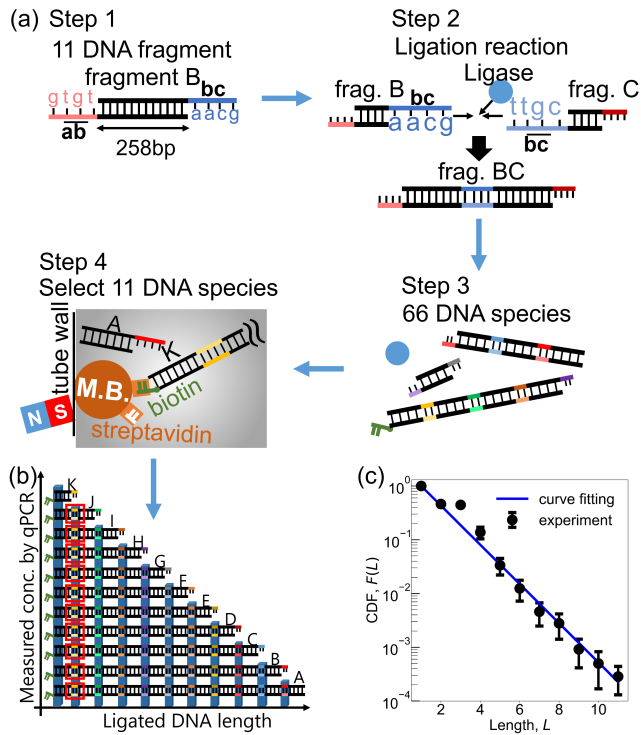}
            \caption{Principle of qPCR-\textit{b}STAR method. (a) Experimental procedure for qPCR-$\textit{b}$STAR. We prepared short DNA fragments (step 1) and performed DNA ligation (step 2). We obtained the 66 species of ligated products (step 3) and then selected subpopulation of 11 DNA species by pull-down method. M.B. stands for magnetic beads. (b) qPCR-\textit{b}STAR quantifies the concentration of ligated DNA products as a cumulative distribution function (CDF) of DNA concentration. By using primers that detect sequences, including jk(red box), we measured the sum of concentrations of \textbf{J}t\textbf{K}, $\cdots$, \textbf{B}t\textbf{K} and \textbf{A}t\textbf{K}. (c) Experimentally measured CDF as a function of DNA length $L$ after ligation in a buffer solution. The experimental data was fitted well with exponential function $\exp(-0.84 \cdot L)$. Error bars represent the standard deviation from three independent experiments. }
            \label{fig1}
        \end{figure}

The experimental protocol of qPCR-\textit{b}STAR is described in Fig. 1(a). 
\begin{itemize}
\item First step: Short DNA fragments are synthesized by PCR amplification (Top left in Fig. 1(a)). These DNA fragments had 11 different sequences and were distinguished by labeling from \textbf{A},  \textbf{B}, $\cdots$, \textbf{K}. Their base pair (bp) number were all the same size and are 258 bp. To make ligated joint between DNA fragments, their both ends had overhang structure called sticky ends. Recognition site by restriction enzyme shown in letters, $\mathrm{\overline{ab}}$ or bc $\cdots$ ($\mathrm{\overline{ab}}$ is complementary sequence of ab), was added to their ends. Sticky ends were obtained as DNA substrate for subsequent end-to-end ligation. 

\item Second step: DNA ligation was performed (Bottom left in Fig. 1(a)). DNAs with sticky ends were ligated by Taq DNA ligase. For instance, \textbf{B} fragment and \textbf{C} fragment became a single DNA chain \textbf{B}t\textbf{C} after ligation since bc and $\mathrm{\overline{bc}}$ formed a hybridized joint.  

\item Third step: Afterwards, the enzyme ligated DNA fragments in an alphabetical order (bottom right in Fig. 1(a))). Ligation of 11 short DNA fragments produced ligated DNA molecules with 66 different sequences and lengths. The ordered sequence was represented by a simple description as \textbf{X}t\textbf{Y}, meaning that short DNA fragments were connected from the \textbf{X} fragment to the \textbf{Y} fragment. For instance, a DNA that includes \textbf{A}-\textbf{B}-\textbf{C}-\textbf{D} sequence is referred to as \textbf{A}t\textbf{D}. While \textbf{K}-\textbf{K} refers to the \textbf{K} fragment alone. In the following description, DNA length is the number of short DNA fragments contained per molecule.

\item Fourth step: In order to find the probability distribution of the concentration of ligated DNA in its length, DNA fragments having the \textbf{K} sequence at the end was extracted and their concentration was measured by qPCR (Bottom left in Fig. 1(a))). We modified the blunt end of \textbf{K} fragment with biotin and extracted 11 target DNA by the pull-down method, using strong binding to streptavidin. The obtained DNA-magnetic beads were used for qPCR analysis.
\end{itemize}
The concentration of selected DNA products containing \textbf{K} fragment reports the distribution function of ligated DNA concentration. The number of unit DNA fragments $L$ ($1 \leq L \leq 11$) is proportional to the number of ligated joints per molecule (that is $L-1$). The concentration of DNA obtained by qPCR analysis $C(L)$ is the cumulative concentration $\sum_{n=L}^{11}c_{DNA}(n)$. Normalized cumulative distribution function (CDF) of the ligated products $F(L)= C(L)/C(1)$ was used to test whether $F(L)$ decays exponentially with respect to DNA size. CDF $F(L)$ gives an experimental clue about the randomness behind enzymatic ligation (Fig. 1(b)).

\subsection*{Random DNA ligation with crowding effect}
We analyzed the ligation reaction in a buffer solution by using qPCR-\textit{b}STAR method. CDF $F(L)$ was plotted for the DNA size $L$ (Fig. \ref{fig1}(c)) and it showed an exponential decay, $F(L)=\exp(-\alpha L)$. The exponential decay of $F(L)$ reflects that the ligation is a random reaction not depending on DNA length itself. A similar exponential decay was also observed in capillary tube electrophoresis analysis though we could detect DNA fragments only shorter than $L \leq 6$ (Fig. S1)\cite{supplement}.

        \begin{figure*}[tbp]
        \centering
\vspace*{3.5em} 
\vspace*{-3em} 
\hspace*{-1em} 
            \includegraphics[scale=0.20]{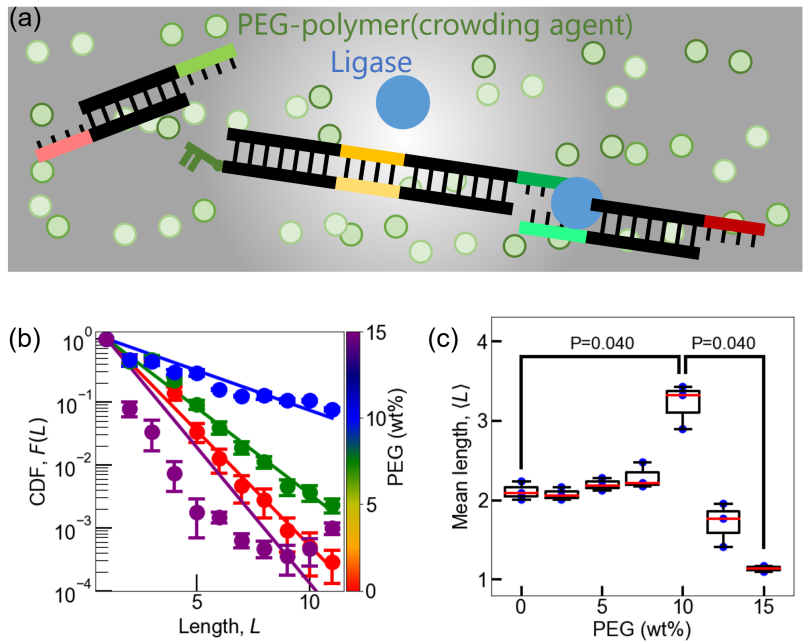}
            \caption{The analysis of DNA ligation in polymer solutions by using qPCR-\textit{b}STAR method. (a) Schematic illustration of DNA ligation in a polymer solution with crowding agent. The green particle represents a crowding agent. We used polyethylene glycol (PEG) as a crowding agent. (b) Experimental result of CDF $F(L)$ of ligated DNA in various PEG concentrations. The color code represents the concentration of crowding agent PEG, from 0\% (red), 7.5\% (green), 10.0\% (blue) to 15.0\% (purple). The fitting curves is $F(L)=\exp(-\alpha L)$ with $\alpha=0.84$ in 0\% PEG, $\alpha=0.63$ in 7.5\% PEG, $\alpha=0.29$ in 10.0\% PEG, $\alpha=0.99$ in 15.0\% PEG. Error bars represent the standard deviation from three independent experiments. (c) Boxplot of mean DNA length $\langle L \rangle$ in PEG concentration. Statistical analysis of Mann-WhitneyU-test was performed. }
            \label{fig2}
        \end{figure*}

Next question is how this exponential decay of $F(L)$ can be suppressed due to physically relevant effects. It has long been studied that molecular crowding of inert polymers increases enzyme activity\cite{zimmerman}, but it is not clear whether ligation reactions occur randomly even in such a crowded environment (Fig. \ref{fig2}(a)). To clarify this point, we performed qPCR-\textit{b}STAR assay for the DNA ligation with various concentration of polyethylene glycol 6000 (PEG) as a crowding agent from 2.5\%to 15.0\% (w/w). As shown in Fig. \ref{fig2}(b), $F(L)$ of the ligated DNA also showed the exponential decay from 2.5\% to 10.0\% PEG, but the slope of exponential decay was more gradual than the decay rate in a buffer solution(0\% PEG). This result means that the ligation reaction occurred at random even in the crowded environment but the number of reaction products was much increased. Moreover, as PEG concentration reached 15.0\%, $F(L)$ showed turned to stretched exponential decay where the tail of the distribution function was stretched slightly (Fig. 2(b), purple). It suggests that DNA ligation has optimal balance of enzymatic reaction and molecular crowding, around at 10.0\% PEG, for efficient synthesis of the longest product.

According to the exponential distribution function $F(L)$ in various PEG concentrations, the mean DNA length (the average number of ligations per molecule) $\langle L \rangle$ was analyzed. Fig. \ref{fig2}(c) shows that $\langle L \rangle$ was increased by the addition of PEG and had a peak at 10.0\% PEG, but it sharply decreased at higher concentrations. This optimal concentration of PEG implies that the molecular crowding enhances the synthesis of longer DNA but the effect underlying this enhanced ligation was gradually suppressed at the over-crowded environment.

\subsection*{Kinetics of DNA ligation}
The qPCR-\textit{b}STAR analysis allows us to find that the DNA ligation is random both in the buffer solution and the crowded environments, but the mean length $\langle L \rangle$ can be affected by the concentration of PEG. However, why is there optimal concentration of PEG at enhanced DNA ligation? To reveal its underlying mechanism, we measured the kinetic increase of ligated products with two pairs of \textbf{K} fragment and another DNA that can be jointed to \textbf{K} fragment. 

We firstly chose the \textbf{J} fragment of same length 258 bp as the short DNA substrate. The rate of the enzymatic reaction in the presence of coexisting PEG was evaluated by measuring the ligated product of the \textbf{K} and \textbf{J} fragments. The ligation reaction started in an aqueous solution (0\% PEG), and a small fraction of solution was taken every 1 min and the concentration of the product \textbf{J}t\textbf{K} was measured by qPCR (Fig. 3(a)). The concentration of \textbf{J}t\textbf{K} increased linearly with time, and the slope of reaction kinetics increased as the concentration of \textbf{J} (Fig. 3(a)). Such linear increase with time was also observed even when PEG 10\% was added (Fig. 3(b)). However, when the PEG concentration exceeded 15\%, although the product \textbf{J}t\textbf{K} increases linearly over time, the slope of reaction curve became independent of the initial concentration of \textbf{J} (Fig. 3(c)). The ligation proceeded at a constant rate regardless of the DNA concentration at 15\% PEG, indicating that ligation reaction is no longer dependent on the number of substrate. 

        \begin{figure*}[tbp]
        \centering
\vspace*{5.5em} 
\vspace*{-5em} 
\hspace*{-1em} 
            \includegraphics[scale=0.21]{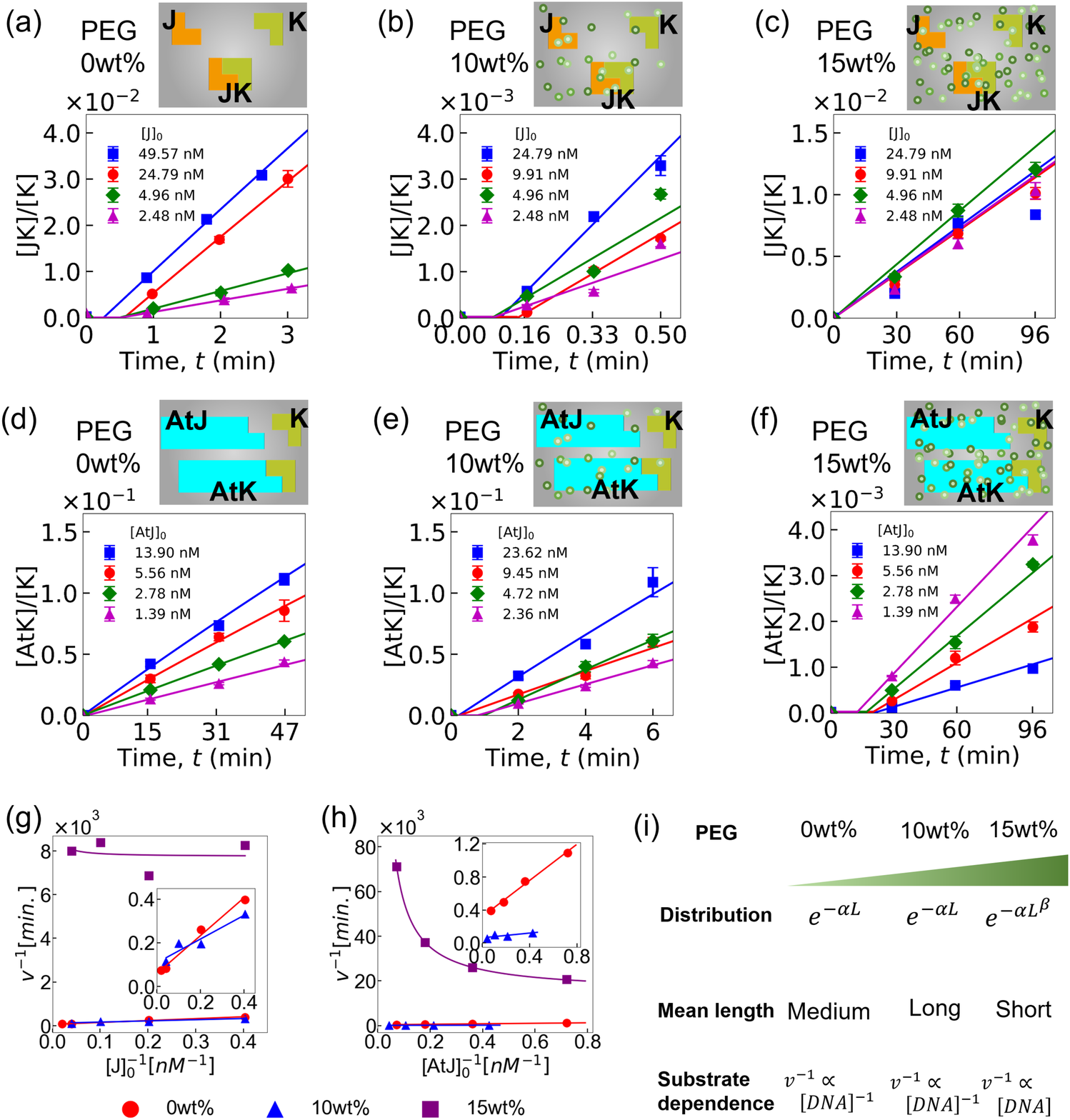}
            \caption{Kinetics of DNA ligation in crowding solutions is analyzed by qPCR-\textit{b}STAR method. (a to c) The relative amount of \textbf{J}t\textbf{K}, which was synthesized from ligation of \textbf{J} and \textbf{K}, was plotted in time $t$ at various PEG concentrations. Color represents the initial concentration of DNA substrate \textbf{J}. (a) 0.0\% PEG, (b) 10.0\% PEG, (c) 15.0\% PEG. (d) 0.0\% PEG, (e) 10.0\% PEG, (f) 15.0\% PEG.  (d to f) The relative amount of \textbf{A}t\textbf{K}, which was synthesized from ligation of \textbf{A}t\textbf{J} and \textbf{K}, was plotted in time $t$ at various PEG concentrations. Color represents the initial concentration of DNA fragment \textbf{A}t\textbf{J}. (g and h) Lineweaver-Burk plot of ligation reactions. (g) DNA ligation of \textbf{J} and \textbf{K} fragments. Fitting curve is linear function. (h) DNA ligation of short \textbf{K} fragment and long \textbf{A}t\textbf{J} fragment. Fitting curve is linear function at 0.0\% and 10.0\% PEG, but it becomes inverse linear function at 15.0\% PEG. (i) Summary of the effect of molecular crowding on the statistical distribution and kinetics of DNA ligation. Randomness, mean length $\langle L \rangle$, and the kinetic speed $v$ show distinct profiles in various PEG concentrations (from left to right: dilute, crowded, highly crowded regimes).}
            \label{fig3}
        \end{figure*}

Next, we chose the \textbf{A}t\textbf{J} fragment that is ligated all 10 fragments from \textbf{A} to \textbf{J}, which is 10 times longer than either the \textbf{K} and \textbf{J} fragment. By analyzing the ligation kinetics of the \textbf{K} and \textbf{A}t\textbf{J}, we found the linear increase of ligation product \textbf{A}t\textbf{K} in both 0\% (Fig. 3(d)) and 10.0\% PEG (Fig. 3(e)), and the slope of kinetics was increased with the initial amount of \textbf{A}t\textbf{J} fragment. Interestingly, in 15.0\% PEG, the slope of kinetics was suppressed as the initial amount of \textbf{A}t\textbf{J} fragment increased (Fig. 3(f)). These results indicate that the presence of macromolecules confers a size dependence of the substrate molecules in the DNA ligation reaction (Figs. 3(g)-(i)). 

\subsection*{Kinetic model of DNA ligation with crowding effect}
We propose theoretical model where two effects are taken into account to explain the concentration dependence of the production rate $v$ and the DNA concentrations of \textbf{J} and \textbf{A}t\textbf{J}. Michaelis-Menten model is known to explain the kinetics of enzymatic reaction by assuming enzyme-substrate complex as an intermediate product. We extended this model by considering intermediate products in order to understand the optimal condition for enhanced DNA ligation in crowding conditions.

        \begin{figure*}[tbp]
        \centering
\vspace*{5.5em} 
\vspace*{-5em} 
\hspace*{-1em} 
            \includegraphics[scale=0.16]{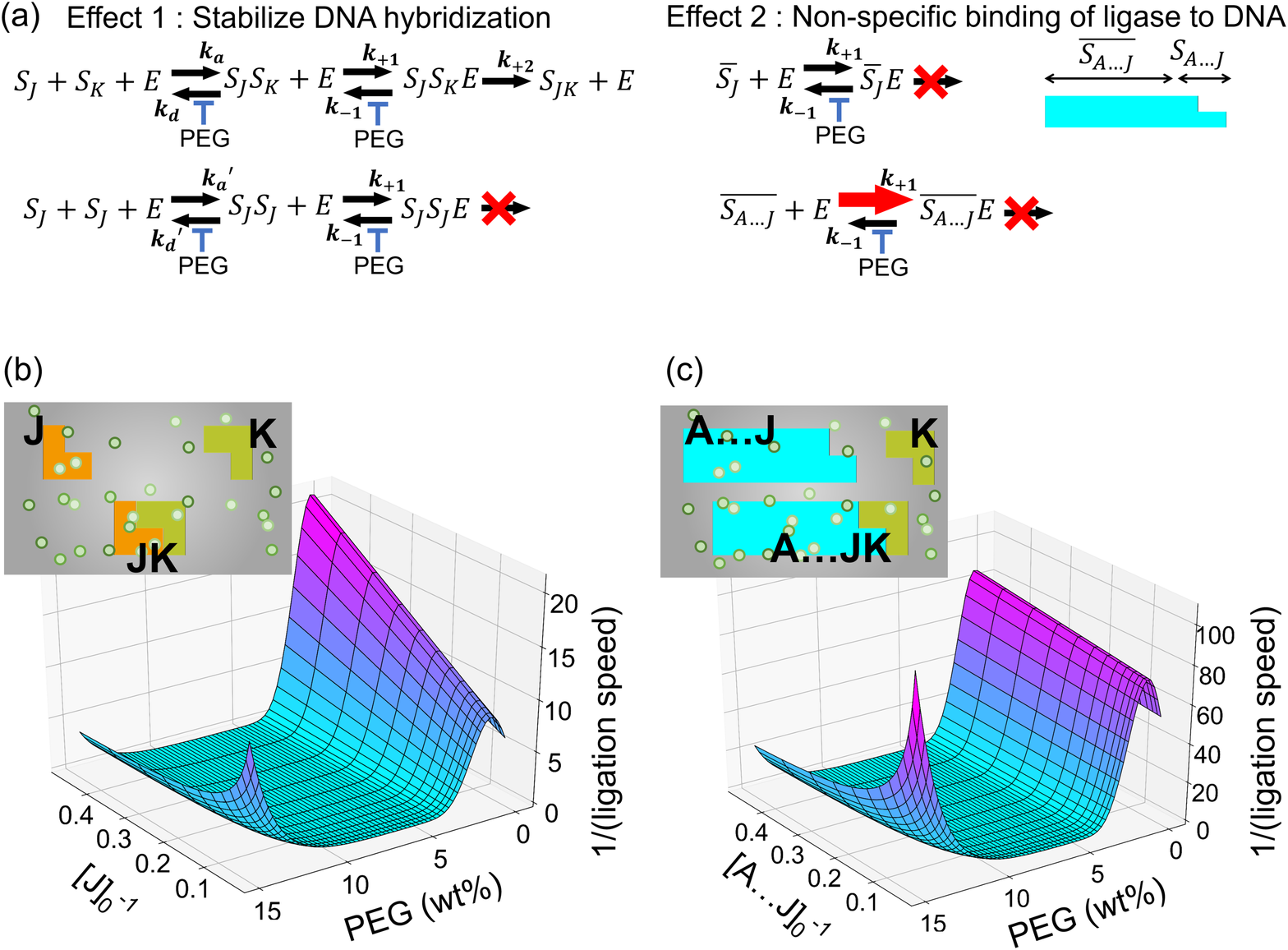}
            \caption{Model of partial hybridization pathways in DNA ligation with molecular crowding. (a) Two distinct effects involved in DNA ligation with molecular crowding. $k_{a}$ is the binding constant of complementary pair, $k_{d}$ is the dissociation constant of hybridized pair, $k_{a}^{\prime}$ is the binding constant of partially complementary pair, $k_{d}^{\prime}$ is the dissociation constant of partially hybridized pair, $k_{+1}$ and $k_{-1}$ are the binding constant and the dissociation constant between DNA and ligase, respectively. $S_{i}^{e}$ is a part of $S_{i}$ including sticky ends. $\overline{S_{i}}$ is a part of $S_{i}$ which do not include sticky ends. 
(b) Theoretical ligation speed for the ligation of two short DNA fragments, \textbf{K} and \textbf{J}. The ligation speed is plotted in the initial DNA concentration $[\textbf{J}]_0$ and the PEG concentration.
(c) Theoretical ligation speed for the ligation of one short DNA fragment \textbf{K} and one long DNA \textbf{A}t\textbf{J}. The ligation speed is plotted in the initial concentration of the long DNA  $[\textbf{A}t\textbf{J}]_0$ and the PEG concentration.
}\label{fig4}
        \end{figure*}

We define [$S_{i}$] as the concentration of DNA $i$ where $i$ stands for one arbitrary sequence from all 66 DNA sequences. The length of DNA $i$ is $L_{i}$. Suppose that ligation reaction mainly consists of three chemical processes: the First process is the hybridization of complementary DNA pair, for instance, \textbf{J} and \textbf{K} fragments. The second step is the complex formation of DNA pair and the ligase enzyme and the third is to trigger the end-to-end DNA ligation at the DNA-ligase complex. To explain the kinetics $v \propto$ [$S_X$]$^{-1}$ as seen in Figs. 3(h) and 3(i), we need to take into account additional pathways to the above three ordinary steps.

One additional effect newly considered is the formation of partial hybridization of a pair of DNA where the sequences of sticky ends are not completely complementary (Fig. 4(a), Effect 1). The sticky ends have 4 base for hybridization, but we assume that the DNA strands can make partially hybridized pairing even in there is one or two base pair mismatch. Although the partial hybridization is weaker in a buffer solution, molecular crowding due to coexisting polymer stabilizes the partial hybridization. 

Another pathway newly considered is that the non-specific interaction of DNA and the ligase. A ligase finds hybridized sticky ends and makes new join between two DNA fragments. We assume that the ligase protein is trapped onto DNA fragment in sequence-independent manner due to the molecular crowding. In dense PEG solution, molecular crowding induces the attractive interaction between the ligase (known as depletion force\cite{marenduzzo}\cite{asakura}). The longer DNA has more space to attract ligase proteins and traps the active enzyme away from the sticky ends (Fig. 4(a), Effect 2). This nonspecific binding onto DNA reduces the freely available enzyme in a solution, meaning that the effective concentration of ligase is decreased in the presence of long DNA in dense PEG solution. These pathways do not contribute to make ligated products but change the reaction diagram, at end-to-end ligation.
By taking into account these pathways appeared in crowded environment, the speed of ligation reaction with two DNA fragments ($S_i$ and $S_j$) is given by,
\begin{eqnarray}
    v_{ij} 
    = \frac{
        k_{+2}[E_{0}][S_{i}][S_{j}]
        }{
            \Psi(c_{peg}) \Theta(\bm{S},c_{peg};\bm{L})
            + \sum_{\substack (l,m)\in{P}}[S_{l}][S_{m}]
        }      
\end{eqnarray}
where $\bm{S}$ represents the concentrations of all 66 DNA ([$S_1$], [$S_2$], $\cdots$ [$S_{66}$]), [$E_0$] is the total concentration of ligase, $c_{peg}$ is the concentration of PEG. $\bm{L}$ represents the lengths of all 66 DNA ($L_1$, $L_2$, $\cdots$, $L_{66}$). $\mathbb{P}$ is a set of all possible DNA pair whose sequence of sticky ends are complementary. The reaction constant of ligation $k_{+2}$ decreases exponentially as a function of $c_{peg}$\cite{supplement}\cite{homchaudhuri}. $\Psi(c_{peg})$ is the dissociation constant among freely available ligase $[E]$, [$S_i$] and [$S_j$], and $\Theta(\bm{S},c_{peg};\bm{L})$ represents the ratio of ligase trapped on DNA and the freely available ligase\cite{supplement}.

We performed the numerical calculation of $v_{ij}$ at various PEG concentration $c_{peg}$ by changing the concentration of $[S_i]$ while keeping $[S_j]$. Experimental results of Figs. 3(g) and 3(h) proposes two cases of crowded ligation as model systems: First one is the ligation of two short DNA fragments $i$ and $j$ that correspond to \textbf{J} and \textbf{K} fragments, and second one is the ligation of long DNA and short DNA that correspond to\textbf{A}t\textbf{J} and \textbf{K} fragments. Fig. 4(b) (and Fig. 4(c)) shows the inverse speed of ligation $v_{\textbf{J}\textbf{K}}^{-1}$ with \textbf{J} fragment (and $v_{\textbf{A}t\textbf{J}\textbf{K}}^{-1}$, respectively) and the concentration of PEG. We find three distinct regime in both surface plots: First, at the buffer solution ($c_{peg}=0$), $v^{-1}$ proportionally increased with [$S_{J}$]$^{-1}$ (and [$S_{\textbf{A}t\textbf{J}}$]$^{-1}$). Second, $v^{-1}$ has local minimum but almost constant even by changing [$S_{J}$]$^{-1}$ (and [$S_{\textbf{A}t\textbf{J}}$]$^{-1}$) at intermediate concentration of PEG such as 10\%,. Finally, $v^{-1}$ slows down as [$S_{J}$]$^{-1}$ increases (and [$S_{\textbf{A}t\textbf{J}}$]$^{-1}$ increases) at highly large concentration of PEG. In particular, the reaction of long and short DNA in Fig, 4(c) clearly shows that the production of longer ligated DNA is suppressed by the addition of excess amount of substrate DNA, as consistent with experimental result. 

These distinct regimes can be explained from Eq.(1) as follows: Given the ligation of short fragments $\bm{S}=([S_{K}], [S_{J}], [S_{JtK}])$ and  $\mathbb{P}$ = (\textbf{J},\textbf{K}), and the ligation of short and long fragments $\bm{S}=([S_{K}], [S_{AtJ}], [S_{AtK}])$ and $\mathbb{P}$ = (\textbf{A}t\textbf{J},\textbf{K}). 

\begin{itemize}
\item \textit{DNA ligation in a buffer solution} 

At a dilute solution of small $c_{peg}$, the effect of molecular crowding is absent. DNA do not contribute to trap ligase enzyme from the solution and $\Theta \sim 1$ and $\Psi/[S_K]$=const. By considering the DNA concentration dependence on ligation speed Eq.(1), $v^{-1} \sim 1+ \frac{\Psi}{[S_K][S_J]} \sim [S_{J}]^{-1}$ (and $[S_{AtJ}]^{-1}$). 

\item \textit{Ligation of two short DNA in a polymer solution}

The molecular crowding (effect 2 in Fig. 4(a)) induces the non-specific binding of ligase onto DNA at the PEG solution. This non-specific binding decreases the freely available enzyme from the solution, and the ratio of trapped enzyme becomes $\Theta \sim [S_{J}][S_{K}]$. By taking into account this effect, the ligation of two short fragments proceeds at $v^{-1} \sim 1+ {\Psi} \sim const$. 

\item \textit{Ligation of long and short DNA in a polymer solution }

Long DNA also traps larger amount of ligase by non-specific binding $\Theta \sim [S_{AtJ}]^{2}$ due to crowding effect (effect 2 in Fig. 4(a)). Hence the ligation of one short fragment and one long fragment is $v^{-1} \sim 1+ \frac{\Psi [S_{AtJ}]^2}{[S_{K}][S_{AtJ}]} \sim [S_{AtJ}]$.
\end{itemize}

This mathematical analysis suggests that the attractive but non-specific binding due to molecular crowding gives rise to the transition of ligation speeds with non-trivial length dependence.

        \begin{figure}[tbp]
        \centering
\vspace*{5.5em} 
\vspace*{-5em} 
\hspace*{-1em} 
            \includegraphics[scale=0.16]{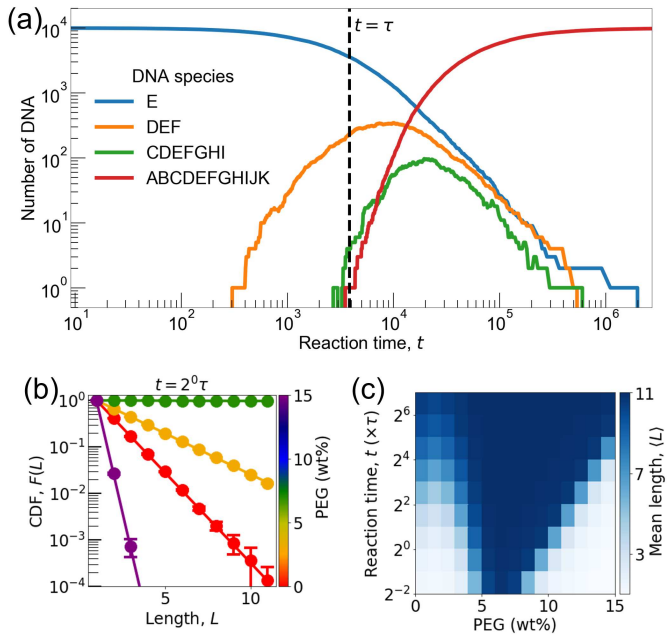}
            \caption{Numerical analysis of stochastic ligation of 11 DNA species in crowded solutions. (a) Ligation kinetics of the number of ligated DNA products. Kinetics of four DNA species are plotted over time. We define the characteristic time $\tau$ when the first longest DNA \textbf{A}t\textbf{K} was synthesized. (b) Normalized cumulative distribution function $F(L)$ with various concentrations of PEG. The color code represents the concentration of crowding agent. The fitting curves is $F(L)=\exp(-\alpha L)$ with $\alpha=0.893$ in 0.0\% PEG, $\alpha=0.407$ in 3.0\% PEG, $\alpha=0.003$ in 7.0\% PEG, $\alpha=3.662$ in 15.0\% PEG. (c) The mean length $\langle L \rangle$ in the concentration of crowding agent and the reaction time $t$ (in unit of $\tau$). The blue color represents the longer mean length.}
            \label{fig5}
        \end{figure}

\subsection*{Optimal crowding effect for long DNA synthesis}
Finally, using the theoretical model based on Eq.(1), numerical simulations were carried out to verify the concentration distribution and the maximization of the long DNA concentration by ligation of a large number of DNA strands.
The ligation kinetics is described by the interplay of the synthesis and the loss of [$S_i$],
\begin{eqnarray}
    \frac{d[S_{i}]}{dt} 
= \sum_{\substack{S_{n}+S_{m}\to S_{i}}} v_{nm} 
- \sum_{(i,l)\in \mathbb{P}} v_{il} .
\end{eqnarray}
The first term represents the production of $S_i$ from all possible DNA pairing (the ligation of $S_{n}+S_{m}\to S_{i}$) and the second term is the loss as substrates for other DNA reactions (the ligation of $S_i+S_l\to S_{i+l}$).
 The joining reaction of 11 kinds of DNA strands was numerically calculated using the Gillespie method, and the concentration distribution at the time when $T$ progressed for a certain time was plotted against the DNA strand length (Fig. 5(a)). As the concentration of PEG was increased, the slope of $F(L)$ became smaller but, as the PEG concentration further increased, the slope of $F(L)$ became steep again. Moreover, this model reproduced that the concentration of longest DNA (Fig. 5(b)) and the mean length of ligated DNA $\langle L \rangle$ (Fig. 5(c)) also had a peak at intermediate PEG concentrations. On the one hand, cCrowding effect promotes the joint formation and increases the rate of ligation per unit time. On the other hand, because the large fraction of ligase would be trapped onto DNA at the higher the PEG concentration, the effective concentration of freely available enzyme becomes lower and in turn the ligation reaction slows down. Such interplay of enhanced binding and reduced enzyme concentration gives rise to optimality for the production of long DNA.

\section*{Discussion}
In this study, we report the enhanced ligation of DNA in crowded polymer solution by using qPCR-based statistical analysis.Heretofore, as a scenario for avoiding the exponential decay due to random ligation, it has been discussed that local concentration amplification cancels the decay by raising the initial concentration\cite{braun1}\cite{braun2}\cite{maeda1}. To increase the proportion of long DNA strands in a molecular population crowded with various kinds of polymer solute\cite{blokhuis}, we found that molecular crowding by a coexisting polymer enhances the efficiency of the enzymatic reaction, in particular DNA ligation. 

Furthermore, we have found the stretched exponential decay of ligated product at highly crowded condition (15.0\% PEG). The deviation from exponential regime implies non-random manner in ligation kinetics. Long-tail distribution such as power-law\cite{hofmann} is also important to suppress exponential decay. Next question is to find the mechanism of enhanced synthesis of long DNA with non-random manner in simple physicochemical condition. It is worthy to note that the microscopic attraction between ligase and DNA is not experimentally detected in this study though it is one plausible mechanism to explain the optimal DNA ligation. The interaction of DNA-binding protein onto DNA has been extensively discussed\cite{elf} and the analysis in detail will be addressed in future study.
Finally, there are crowded conditions in which high concentrations of proteins are present in living cells\cite{poolman}. Whether enzymes react randomly or non-randomly in these circumstances is little understood. One of challenges for future studies is to link the dynamics of one enzyme molecule in a living cell with appearing functions in the whole intracellular environment.

 \subsection*{Materials and Methods}
\subsubsection{Synthesis of short DNA fragments}
Short DNA fragments used as substrates for ligation reactions were synthesized through polymerase chain reaction (PCR). DNA from the pUC19 plasmid was used as template for the synthesis of 11 different non-overlapping DNA fragments, and a KOD-plus-Neo polymerase (KOD-401, Toyobo) was used for DNA amplification. Resulting DNA fragments were 258-bp long and were labeled alphabetically as \textbf{A}, \textbf{B}, $\cdots$, \textbf{K}. Oligonucleotide primers were designed by using Primer3web\cite{Untergasser}\cite{Koressaar1}\cite{Koressaar2} and were purchased from Eurofins Genomics. The sequences of oligo primers used in this study are listed in the Supplementary table\cite{supplement}.

Synthesized DNA was treated with a restriction enzyme with \textit{Bst}X1 (R0113L, New England BioLabs) at \SI{37.0}{\celsius} overnight. \textit{Bst}X1 recognition sites were added to the $5^{\prime}$ end of each primer. Purified DNA fragments have a sticky end at each side, except those synthesized using the forward primer for the \textbf{A} fragment and the reverse primer for the \textbf{K} fragment. In addition, the reverse primer for the \textbf{K} fragment has a biotin in its $5^{\prime}$ end in order to capture $5^{\prime}$ biotinylated \textbf{K} fragment for the pull-down step (Fig.1(a), step4).

\subsubsection{DNA ligation}
The final volume of the ligation reaction solution was \SI{10}{\micro\liter}, where \SI{7.27}{\nano\gram} of 11 short DNA fragments and  thermostable Taq DNA ligase (M0208, New England Biolab) at a final concentration of 20 U were mixed. Ligation reaction was performed at \SI{25.0}{\celsius} for \SI{15}{\hour}. Short DNA fragments were designed to be linked in alphabetical order, i.e. \textbf{A} fragment was linked to \textbf{B} fragment, and \textbf{B} fragment bind to \textbf{C} fragment.

\subsubsection{qPCR analysis}
The qPCR reaction volume was \SI{10}{\micro\liter}. We mixed \SI{5}{\micro\liter} of 2$\times$TB Green Premix Ex Taq II  (RR820, TaKaRa) with \SI{0.25}{\micro\liter} of extracted DNA-magnetic beads, \SI{0.5}{\micro\liter} of the forward primer, \SI{0.5}{\micro\liter} of the reverse primer, and \SI{3.75}{\micro\liter} of MilliQ water. The final concentration for each DNA primer was \SI{400}{\nano M}. Thermal cycling for qPCR analysis was performed using a real-time PCR system (CFX96, Bio-Rad) as follows: heating to \SI{95.0}{\celsius} for 30 s; 35 cycles of \SI{95.0}{\celsius} for \SI{10}{\second} and \SI{60.0}{\celsius} for \SI{30}{\second}. The threshold cycle number $C_q$ was determined and the initial DNA concentration was calculated according with this value as follows: Using pUC19 plasmid DNA as a template DNA of a known concentration, the $C_q$ value of each primer was plotted into a calibration curve between the $C_q$ value and DNA concentration

For the measurement of ligation speed $v_{ij}$, we mixed \textbf{K} fragment (final conc. \SI{48.6}{\pico M}) with either \textbf{J} fragment or \textbf{AtJ} fragment at various concentrations. The ligation reaction was performed at \SI{25.0}{\celsius} with a thermostable Taq DNA ligase (final conc. 6.4 U) and the volume of reaction mixture was \SI{4}{\micro\liter}. We chose time steps to observe linear increment of the relative amount of ligated DNA. \\

\textit{Acknowledgements}

This work was supported by Grant-in-Aid for Scientific Research on Innovative Areas (JP16H00805 Synergy of Structure and Fluctuation, JP17H05234 Hadean Bioscience, and JP18H05427 Molecular Engines) and Grant-in-Aid for Scientific Research (B) JP17KT0025 from MEXT, and Human Frontier Science Program Research Grant (RGP0037/2015). We thank R. Sakai, K. Yoshimoto and S. Terada for experimental assistance.\\

\textit{Competing interests}

The authors declare no competing interests.\\

\textit{Corresponding address}

shiraki@phys.kyushu-u.ac.jp

\newpage

$ $

\newpage

\onecolumngrid

\renewcommand{\figurename}{FIG. S}

\renewcommand{\tablename}{TABLE S}
\renewcommand{\refname}{References}
\renewcommand{\arraystretch}{1.2}
\setcounter{figure}{0}
\title{Supplemental information for \\ Randomness and optimality in enhanced DNA ligation with crowding effects}
\author{Takaharu Shiraki$^1$, Ken-ichiro Kamei$^2$, and Yusuke T. Maeda$^{1}$}
\affiliation{$^1$Kyushu University, Department of Physics, Motooka 744, Fukuoka 812-0395, Japan}
\affiliation{$^2$Kyoto University, Institute for Integrated Cell-Material Sciences (iCeMS), Yoshida-Ushinomiyacho, Sakyo-ku, Kyoto 606-8501, Japan}

\maketitle

\section{Supplemental materials and methods}

\subsection*{Capillary electrophoresis}
We also checked the exponential distribution of DNA concentration with its size by capillary electrophoresis. An electrophoresis band as shown in FIG.S1(a) was obtained using an electrophoresis apparatus (Agilent, Bioanalyzer2100). When the DNA concentration was quantified from the concentration of these bands, the exponential decay was observed (Fig. S1(b)) as same as qPCR-\textit{b}STAR analysis. However, in capillary electrophoresis, a band with a DNA length of 7 or more could not be detected. This result supports the advantage of qPCR-\textit{b}STAR analysis in order to quantitatively measure the distribution function.
        \begin{figure}[tbh]
          \includegraphics[width=12cm]{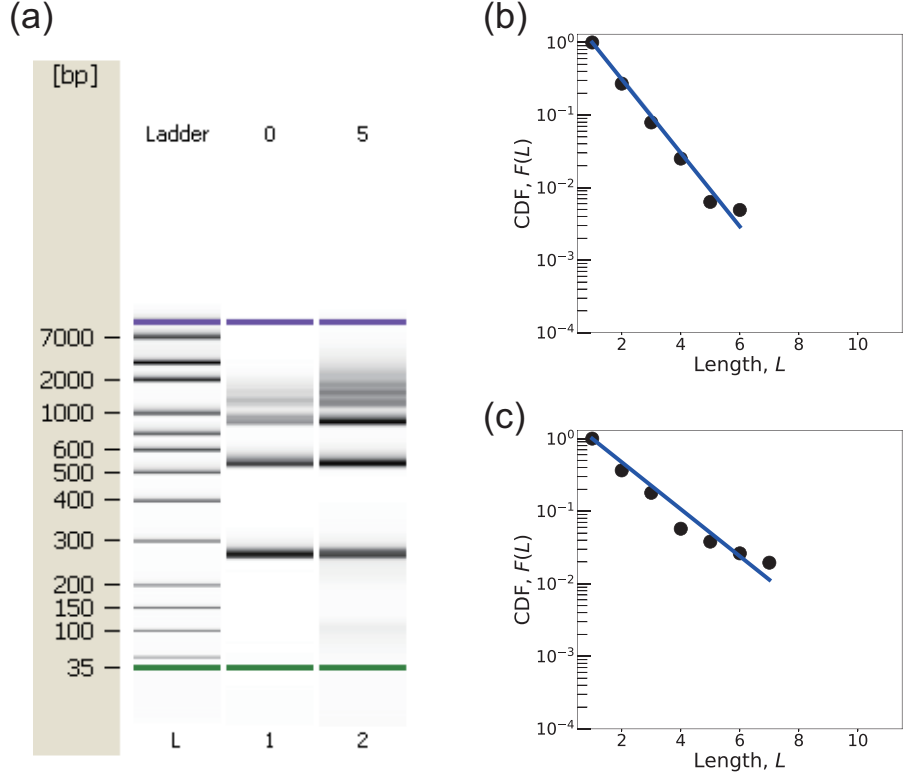}
            \caption{
Capillary electrophoresis assay of ligated DNA. (a) Bands of DNA separated by capillary electrophoresis. Left lane is marker DNA, middle lane is ligated DNA in a buffer solution, right lane is ligated DNA in 5.0\% PEG. (b) Concentration of DNA bands, which are products of DNA ligation in a buffer solution, with respect to the DNA length. Black dots are experimental data and blue solid line is fitting curve of exponential function. (c) Concentration of DNA bands, which are products of DNA ligation in the 5.0\% PEG solution, with respect to the DNA length. Black dots are experimental data and blue solid line is fitting curve of exponential function.}
            \label{figS2} 
        \end{figure}

\subsection*{ qPCR quantification}
\begin{figure}[b]
            \includegraphics[width=12cm]{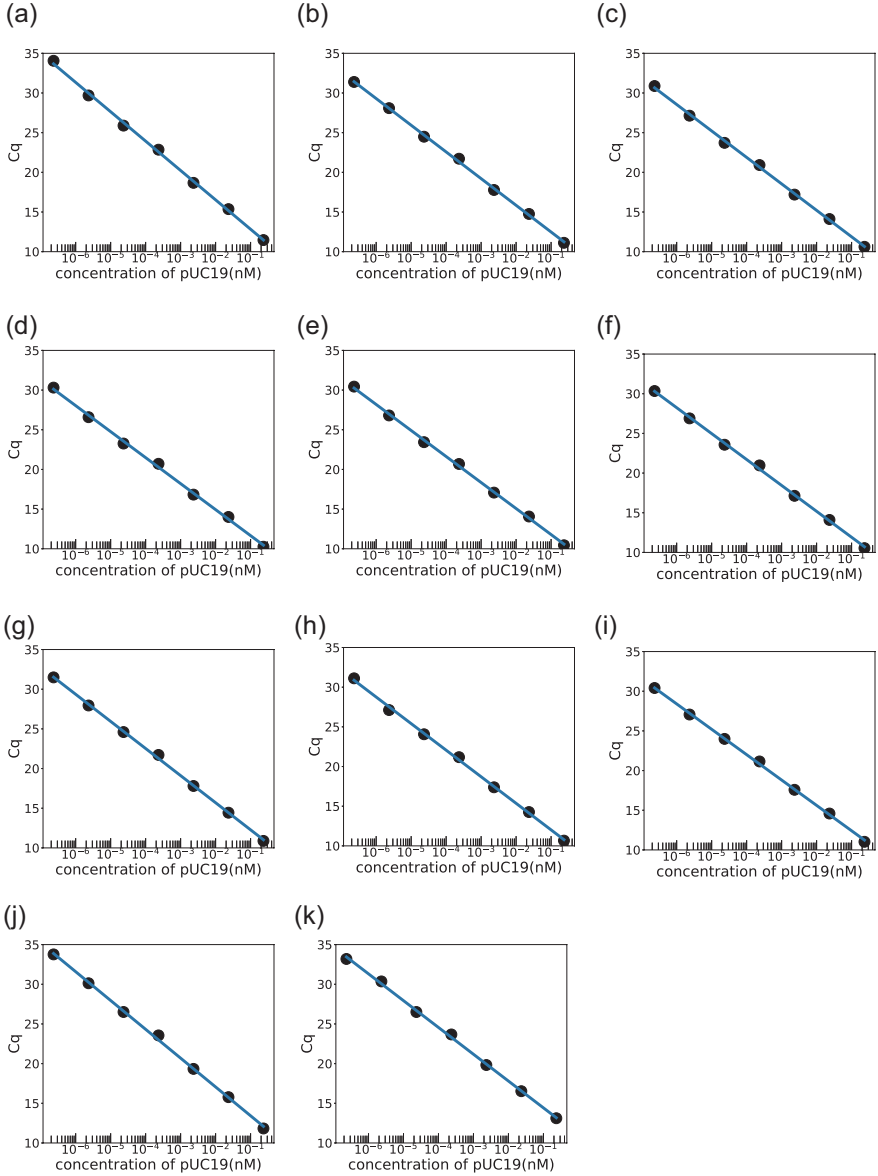}
            \caption{
Calibration curve of threshold cycle $C_q$ and given DNA concentration. (a to k) }
            \label{figS1} 
        \end{figure}

As described in main text, short DNA fragments used in end-to-end ligation were made by polymerase chain reaction (PCR) with KOD-plus-Neo polymerase (KOD-401, Toyobo). We used pUC19 plasmid DNA as template for PCR amplification and 11 different sequences were selected short DNA fragments. Resulting DNA fragments were 258-bp long and were labeled alphabetically as \textbf{A}, \textbf{B}, $\cdots$, \textbf{K}. As desbribed in main text, we represent the ordered sequence of ligated DNA products as \textbf{X}t\textbf{Y}, meaning that short DNA fragments from the \textbf{X} fragment to the \textbf{Y} fragment are connected into one DNA polymer. For instance, a DNA that includes \textbf{A}-\textbf{B}-\textbf{C}-\textbf{D} sequence is referred to as \textbf{A}t\textbf{D}. While \textbf{K}-\textbf{K} refers to the \textbf{K} fragment alone. We designed qPCR primers that can make hybridization pairing at the junctions connecting \textbf{A}-\textbf{B}, \textbf{B}-\textbf{C}, $\cdots$, and \textbf{J}-\textbf{K}. Because these qPCR primers also have complementary sequences to pUC19 plasmd DNA, we can also obtain qPCR growth curve by using pUC19 DNA. This growth curve allows us to draw a calibration curve between the $C_q$ value with each primer set and the actual DNA concentration (Fig. S\ref{figS1}). This experiment was performed with all qPCR primer sets.

\subsection*{Protocol for qPCR-\textit{b}STAR}
\begin{itemize}
\item qPCR-\textit{b}STAR Step 1: DNA ligation.

 The ligation of 11 short DNA fragments (\textbf{A}, \textbf{B}, $\cdots, \textbf{K}$) was performed in a test tube at a temperature of \SI{25.0}{\celsius} for 15 hours (Fig.1(a) in main text, first and second steps). Each short DNA fragment had sticky-ends after digestion with the restriction enzyme \textit{Bst}X1. We designed an ordered paring, for instance \textbf{A} fragment binds to \textbf{B} fragment, and \textbf{B} fragment binds to \textbf{A} fragment at one end and to \textbf{C} fragment at the other end. 

$ $

\item qPCR-\textit{b}STAR Step 2: Selection of DNA fragments for analysis.

The ligation of 11 different short DNA fragments in a single tube produces 66 types of ligated DNA products (Fig.1(a), third step).  Rather than measuring the concentration of each of the 66 types of DNA, we extracted a subpopulation of the 11 types of DNA molecules that have \textbf{K} sequence in common as \textbf{A}t\textbf{K}, \textbf{B}t\textbf{K}, \textbf{C}t\textbf{K}, $\cdots$, \textbf{K}, but its length covers all possible size. For this aim, the \textbf{K} fragment was biotinylated at its 3' end. This modification allowed selecting our target population by using a streptavidin pull-down purification method (Fig. 1(a), fourth step). We utilized magnetic beads covered with streptavidin, Dynabeads M-270 Streptavidin (Thermo Fisher Scientific), that bound tightly to the biotin conjugated to the \textbf{K} fragment, for the separation of our target molecules from the rest. These DNA molecules containing a \textbf{K}-sequence were subjected to subsequent qPCR analysis.

$ $

\item qPCR-\textit{b}STAR Step 3: qPCR analysis.

Then, the concentration of each ligation junction and \textbf{K} fragment was measured (Fig.1(b)). The concentration of DNA detected by qPCR primer set \textbf{K} reported the total concentration of all 11 DNA types; it was defined as $C(1)=\sum_{n=1}^{11}c(n)$ where $n$ is the number of DNA fragments per molecule (Fig.2(a)). In the same way, the concentration of the ligation junction between the \textbf{J} fragment and the \textbf{K} fragment was defined as $C(2)=\sum_{n=2}^{11}c(n)$, which reported the total concentration of DNA molecules that have a \textbf{J}-\textbf{K} junction, i.e.  \textbf{A}t\textbf{K}, \textbf{B}t\textbf{K}, $\cdots$, \textbf{J}t\textbf{K}. A more general definition is that the measured concentration of a ligation junction between the \textbf{X} fragment and the \textbf{X+1} fragment equals the total concentration from \textbf{A}t\textbf{K}, $\cdots$, to \textbf{X}t\textbf{K}. Therefore, the concentration obtained from qPCR analysis is the cumulative concentration of $S(L)=\sum_{n=L}^{11}c(n)$. For the analysis of experimental data, a normalized cumulative distribution function (CDF) $F(L):= C(L)/C(1)=\sum_{n=L}^{11}c(n)/\sum_{n=1}^{11}c(n)$ was used, because the exponential function is retained if the cumulative distribution function has a large $L$. CDF $F(L)$ was also expressed as $F(L) =\sum_{n=L}^{11}p(n)$ by using a probability distribution function  $p(n)=c(n)/\sum_{n=1}^{11}c(n)$.
\end{itemize}

\subsection*{Oligo DNA primer sequence}
Sequences of oligo DNA primer used in the synthesis of 11 short DNA fragments are listed in Supplemental Table I, and sequences of oligo DNA primer for the qPCR quantification are listed in Supplemental Table II. 

\begin{figure}[bth]
            \includegraphics[width=10cm]{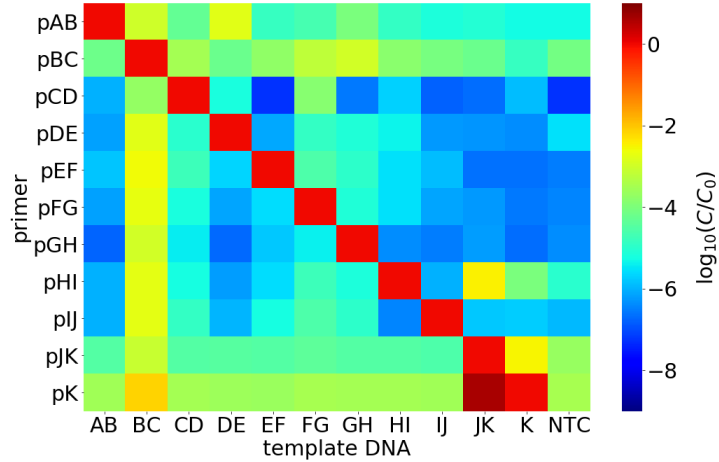}
                \caption{Nonspecific amplification was checked. We measured concentration of each DNA(0.9 pM)(AB,$\cdots$,K) which has only one junction by using primer sets shown in TABLE. S\ref{qpcrset}. $C_{0}$ is concentration of DNA measured by a matched primer set(i.e. concentration of AB measured by pAB), and $C$ is concentration of DNA measured by other primer sets(i.e. concentration of AB measured by pBC). NTC means no template control.}
                \label{fig_qPCRUnique}
\end{figure}

We also confirmed that non-specific detection by qPCR primers is negligible (FIG.S\ref{fig_qPCRUnique}). One primer set is designed to uniquely detect one DNA sequence.  Even if the wrong sequence is read, its error concentration is very small, less than $10^{-2}$ times of the DNA with the correct sequence. This detection error does not affect qPCR-\textit{b}STAR analysis.

\begin{table}[htb]
    \begin{tabular}{|c||c|c|c|} \hline
      fragment 
      & direction 
      & sequence($5^{\prime}-3^{\prime}$) 
      \\ \hline \hline
      A 
      & forward
      & GCGCCCAATACGCAAACCGCCTCTC \\ 
      & reverse
      & \underline{CCCCAA{\bf TGTG}CTGG}CTGCAGGCATGCAA \\ \hline
      B 
      & forward
      & \underline{CCCCAG{\bf CACA}TTGG}GTCGACTCTAGAGGA \\ 
      & reverse
      & \underline{CCCCAA{\bf GCAA}CTGG}TATGCGGTGTGAAATA \\ \hline
      C 
      & forward
      & \underline{CCCCAG{\bf TTGC}TTGG}TGGTGCACTCTCAGTACAATC \\ 
      & reverse
      & \underline{CCCCAA{\bf GGAT}CTGG}TCTAAGAAACCATTATTATCA \\ \hline
      D 
      & forward
      & \underline{CCCCAG{\bf ATCC}TTGG}CGTCAGGTGGCACTTTTC \\ 
      & reverse
      & \underline{CCCCAA{\bf GGTA}CTGG}CCCAACTGATCTTCAGCATC \\ \hline
      E 
      & forward
      & \underline{CCCCAG{\bf TACC}TTGG}TGCACGAGTGGGTTACA \\ 
      & reverse
      & \underline{CCCCAA{\bf ATGC}CTGG}ACTGCATAATTCTCTTACTGT \\ \hline
      F 
      & forward
      & \underline{CCCCAG{\bf GCAT}TTGG}GCTGCCATAACCATGAGTGAT \\ 
      & reverse
      & \underline{CCCCAA{\bf ATCG}CTGG}TTAATTGTTGCCGGGAAGCTA \\ \hline
      G 
      & forward
      & \underline{CCCCAG{\bf CGAT}TTGG}TAGACTGGATGGAGGCGGATA \\ 
      & reverse
      & \underline{CCCCAA{\bf TAGC}CTGG}GACAGTTACCAATGCTTAATC \\ \hline
      H 
      & forward
      & \underline{CCCCAG{\bf GCTA}TTGG}AGACCAAGTTTACTCATATAT \\ 
      & reverse
      & \underline{CCCCAA{\bf TACG}CTGG}TTGATCCGGCAAACAAACCAC \\ \hline
      I 
      & forward
      & \underline{CCCCAG{\bf CGTA}TTGG}GAGCTACCAACTCTTTTTCCG \\ 
      & reverse
      & \underline{CCCCAA{\bf TGCT}CTGG}TGCACGAACCCCCCGTTCAGC \\ \hline
      J 
      & forward
      & \underline{CCCCAG{\bf AGCA}TTGG}CACAGCCCAGCTTGGAGCGAA \\ 
      & reverse
      & \underline{CCCCAA{\bf ACGA}CTGG}CATAGGCTCCGCCCCCCTGAC \\ \hline
      K 
      & forward
      & \underline{CCCCAG{\bf TCGT}TTGG}GAAAAACGCCAGCAACGCGGC \\ 
      & reverse
      & [BioON]GTCGTGCCAGCTGCATTAAT \\ \hline
    \end{tabular}
    \caption{The list of oligo primers used for synthesis of short DNA fragments by PCR and their sequences. Sticky ends cut by restriction enzyme treatment are shown in bold. The target sequence recognized by restriction enzyme are underlined.}
\end{table}
$ $

\newpage

\begin{table}[hbt]
    \begin{tabular}{|c|c||c|c|c|} \hline
        name
        & recognized junction 
        & direction 
        & sequence($5^\prime-3^\prime$) 
        \\ \hline \hline
        pAB
        & A-B 
        & forward
        & TCCGGCTCGTATGTTGTGTG \\ 
        &
        & reverse
        & TTGTAAAACGACGGCCAGTG \\ \hline
        pBC
        & B-C 
        & forward
        & CGCATCTGTGCGGTATTTCAC \\ 
        &
        & reverse
        & GCTTGTCTGTAAGCGGATGC \\ \hline
        pCD
        & C-D 
        & forward
        & ACGAAAGGGCCTCGTGATAC \\ 
        &
        & reverse
        & GTCTCATGAGCGGATACATATTTGA \\ \hline
        pDE
        & D-E 
        & forward
        & TTTCCGTGTCGCCCTTATTC \\ 
        &
        & reverse
        & GGGCGAAAACTCTCAAGGA \\ \hline
        pEF
        & E-F 
        & forward
        & TCGCCGCATACACTATTCTCA \\ 
        &
        & reverse
        & GAAGTAAGTTGGCCGCAGTG \\ \hline
        pFG
        & F-G 
        & forward
        & CGGAGCTGAATGAAGCCATAC \\ 
        &
        & reverse
        & ACTTTATCCGCCTCCATCCA \\ \hline
        pGH
        & G-H 
        & forward
        & CGAAATAGACAGATCGCTGAGATAG \\ 
        &
        & reverse
        & ACTCACGTTAAGGGATTTTGGTCA \\ \hline
        pHI
        & H-I 
        & forward
        & GCGTCAGACCCCGTAGAAA \\ 
        &
        & reverse
        & GCCAGTTACCTTCGGAAAAAGA \\ \hline
        pIJ
        & I-J 
        & forward
        & AGTCGTGTCTTACCGGGTTG \\ 
        &
        & reverse
        & TGGCGCTTTCTCATAGCTCA \\ \hline
        pJK
        & J-K 
        & forward
        & TTCGCCACCTCTGACTTGA \\ 
        &
        & reverse
        & GCAGGAAAGAACATGTGAGCA \\ \hline
        pK
        & K 
        & forward
        & CTTTTGCTGGCCTTTTGCTC \\ 
        &
        & reverse
        & CTTCCTCGCTCACTGACTCG \\ \hline
      \end{tabular}
    \caption{The list of oligo DNA primers used for qPCR analysis and their sequences.}\label{qpcrset}
\end{table}

$ $

\newpage

\begin{table}[tb]
    \begin{tabular}{|c||c|c|c|} \hline
        fragment 
        & direction 
        & sequence($5^\prime-3^\prime$) 
        \\ \hline \hline
        AB 
        & forward
        & GCGCCCAATACGCAAACCGCCTCTC \\ 
        & reverse
        & CCCCAAGCAACTGGTATGCGGTGTGAAATA \\ \hline
        BC 
        & forward
        & CCCCAGCACATTGGGTCGACTCTAGAGGA \\ 
        & reverse
        & CCCCAAGGATCTGGTCTAAGAAACCATTATTATCA \\ \hline
        CD 
        & forward
        & CCCCAGTTGCTTGGTGGTGCACTCTCAGTACAATC \\ 
        & reverse
        & CCCCAAGGTACTGGCCCAACTGATCTTCAGCATC \\ \hline
        DE 
        & forward
        & CCCCAGATCCTTGGCGTCAGGTGGCACTTTTC \\ 
        & reverse
        & CCCCAAATGCCTGGACTGCATAATTCTCTTACTGT \\ \hline
        EF 
        & forward
        & CCCCAGTACCTTGGTGCACGAGTGGGTTACA \\ 
        & reverse
        & CCCCAAATCGCTGGTTAATTGTTGCCGGGAAGCTA \\ \hline
        FG 
        & forward
        & CCCCAGGCATTTGGGCTGCCATAACCATGAGTGAT \\ 
        & reverse
        & CCCCAATAGCCTGGGACAGTTACCAATGCTTAATC \\ \hline
        GH 
        & forward
        & CCCCAGCGATTTGGTAGACTGGATGGAGGCGGATA \\ 
        & reverse
        & CCCCAATACGCTGGTTGATCCGGCAAACAAACCAC \\ \hline
        HI 
        & forward
        & CCCCAGGCTATTGGAGACCAAGTTTACTCATATAT \\ 
        & reverse
        & CCCCAATGCTCTGGTGCACGAACCCCCCGTTCAGC \\ \hline
        IJ 
        & forward
        & CCCCAGCGTATTGGGAGCTACCAACTCTTTTTCCG \\ 
        & reverse
        & CCCCAAACGACTGGCATAGGCTCCGCCCCCCTGAC \\ \hline
        JK 
        & forward
        & CCCCAGAGCATTGGCACAGCCCAGCTTGGAGCGAA \\ 
        & reverse
        & [BioON]GTCGTGCCAGCTGCATTAAT \\ \hline
        K 
        & forward
        & CCCCAGTCGTTTGGGAAAAACGCCAGCAACGCGGC \\ 
        & reverse
        & [BioON]GTCGTGCCAGCTGCATTAAT \\ \hline
      \end{tabular}
    \caption{The list of oligo DNA primers used for the error-estimation in qPCR measurement.The alphabet set listed in the left column is the label sequence (\textbf{AB}, \textbf{BC}, $\cdots$, \textbf{JK}, \textbf{K}) of the synthesized DNA. We used these DNA for the analysis of uniqueness of the sets of qPCR primer. The experimental result is shown in FIG.S\ref{fig_qPCRUnique}}
    \label{primerForUniqueTest}
\end{table}

$ $

\newpage

\setcounter{equation}{0}

$ $

\newpage

\section{Theoretical model} 
\subsection{Kinetic model for DNA ligation in crowded solution}

In this section, we show the derivation of Eq.(1) in maintext where the speed of ligation in crowded condition is given by the concentration of DNA fragments with sticky ends. First, we 
\begin{itemize}
    \item $E$ : ligase
    \item $S_{i}$ : DNA $i$  
    \item $[S_{i}]$ : concentration of DNA $i$ (mol/L)
    \item $L_{i}$ : length of DNA $i$ (bp)
    \item $l_{L}$ : length of DNA where ligase search ligation site while ligase attach to DNA (bp)
    \item $S_{i}S_{j}$ : complex of DNA $i$ and DNA $j$ by complementary base-pairing interactions between adhered cohesive ends
    
    \item $ES_{i}S_{j}$ : complex of ligase and $S_{i}S_{j}$
    \item $S_{i,j}$ : DNA which is ligated DNA $i$ and DNA $j$
\item $L_{i,j} = L_{i}+ L_{j}$
    \item $\mathbb{P} = \{(i,j)|$ sticky end of DNA i and that of DNA j is complementary.$\}$
    \item $\mathbb{I} = \{(i,j)|$sticky end of DNA i and that of DNA j is NOT complementary.$\}$
\end{itemize}

Consider $N$ species of DNA and a ligation reaction of $S_{i} + S_{j} \to S_{k}$.
Because the short homologous DNA sequences make hybridization of double strand DNA structure at the sticky ends, DNA ligase recognizes this hybridized DNA region and subsequently carries out homologous recombination, thus DNA can be defined as two distinct regions:
\begin{itemize}
    \item $S_{i}^{e}$ : homologous $l_{L}$(bp) sequence at sticky ends of DNA $S_{i}$ which sticks to DNA $S_{j}$.
    \item $\overline{S_{i}}$ : non-homologous area of DNA $S_{i}$
\end{itemize}
Because $S_{i}$ has only 1 $S_{i}^{e}$, the concentrations of $S_{i}^{e}$ and $S_{i}$ holds the following conservation rule
\begin{equation}
	\label{defOfSe}
    [S_{i}^{e}] = [S_{i}].
\end{equation}
On the other hand, the length of $\overline{S_{i}}$ is $L_{i}-l_{L}$ and the concentration of $\overline{S_{i}}$ can be described as follows
\begin{equation}
    [\overline{S_{i}}] = M(S_{i};l_{L}) [S_{i}]   .
\end{equation}
with prefactor $M(S_{i};l_{L})$ is given by
\begin{equation}
    M(S_{i};l_{L}) \equiv \left\{ \begin{array}{ll}
        L_{i}/l_{L} - 1 & (L_{i}/l_{L} \geq 1) \\
        0 & (L_{i}/l_{L} < 1)   ,
      \end{array} \right.
\end{equation}

$ $ 

By considering these distinct DNA sequences shown above, the DNA ligation reaction is described by the following chemical reactions($(i,j) \in \mathbb{P}$);
Hybridization and dissociation of $S_{i}$ and $S_{j}$ are given by
\begin{eqnarray}
	\label{normalHybDis}
    S_{i} + S_{j} \ \mathop{\rightleftharpoons}_{k_{d}}^{k_a} \ S_{i}S_{j} .
\end{eqnarray}

The kinetic constant of association and dissociation of ligase on the hybridized sticky ends is given by
\begin{eqnarray}
	\label{FormEplusSiSj}
    E + (S_{i}S_{j})^{e} \ \mathop{\rightleftharpoons}_{k_{-1}}^{k_{+1}} \ E(S_{i}S_{j})^{e} .
\end{eqnarray}
The binding and dissociation of ligase at homologous sequence of $S_{i}$ that can make hybridization pairing with $S_{j}$ and vice versa is given by
\begin{eqnarray}
	\label{FormESie}
    E + S_{i}^{e} \ \mathop{\rightleftharpoons}_{k_{-1}}^{k_{+1}} \ ES_{i}^{e} .
\end{eqnarray}
Hybridization of $S_{j(i)}$ and $S_{i(j)}$ which has $E$ on its homologous sequence are given by
\begin{eqnarray}
	\label{ESiplusSj}
    ES_{i(j)}^{e} + S_{j(i)} \ \mathop{\rightleftharpoons}_{k_{d}}^{k_a} \ E(S_{i}S_{j})^{e} .
\end{eqnarray}
The binding and dissociation of ligase at non-homologous sequence of $S_{i(j)}$ is given by
\begin{eqnarray}
	\label{nonHomologous}
    E + \overline{S_{i(j)}} \ \mathop{\rightleftharpoons}_{k_{-1}}^{k_{+1}} \ E\overline{S_{i(j)}} .
\end{eqnarray}
Finally, the synthesis of ligated product is is given by
\begin{eqnarray}
	\label{endLigation}
    E(S_{i}S_{j})^{e} \xrightarrow{k_{+2}} E + S_{k}   .
\end{eqnarray}
Chemical reactions shown above are limited to the reaction pathway among two DNA substrate with perfect base pairing in hybridization sequence, but in order to explain the optimal enhanced DNA ligation in crowded solution, we additionally assume the imperfect hybridization of two DNA $S_{l}$ and $S_{m}$ where their sticky ends do not completely match as complementary DNA sequence ($(l,m) \in \mathbb{I}$);
\begin{eqnarray}
	\label{mismatchHyb}
    S_{l} + S_{m} \ \mathop{\rightleftharpoons}_{k_{d^{\prime}}}^{k_{a^{\prime}}} \ S_{l}S_{m} .
\end{eqnarray}
Such partial hybridization can be induced by the presence of crowding agent given that the attractive interaction in large polymer arises from the depletion force in crowded polymer solution.

\section{Theoretical analysis of ligation speed in crowded environment}

\subsection*{Setup for chemical reaction}
Next, we derive the speed of ligation of $S_{i}+S_{j}$  $v_{ij}$ $((i,j)\in\mathbb{P})$ by solving Michaelis-Menten equation of the chemical reactions Eqs. (\ref{normalHybDis}) to (\ref{mismatchHyb}).

The kinetics of the complex synthesis $[E(S_{i}S_{j})^{e}]$ is 
\begin{eqnarray}\label{lig1}
    \frac{\mathrm{d}}{\mathrm{d} t} [E(S_{i}S_{j})^{e}] = 
    k_{+1}[E][(S_{i}S_{j})^{e}] + k_{a}([ES_{i}^{e}][S_{j}^{e}] + [ES_{j}^{e}][S_{i}^{e}])
    - (k_{-1} + 2 k_{d} + k_{+2})[E(S_{i}S_{j})^{e}]   
\end{eqnarray}
where $(i,j)\in\mathbb{P}$.
We have to convert $[E(S_{i}S_{j})^{e}]$ to $C[E][S_{i}][S_{j}]$ with constant $C$. 
At the steady-state $\frac{\mathrm{d}}{\mathrm{d} t} [E(S_{i}S_{j})^{e}]=0$, Eq. (\ref{lig1}) is rewritten by 
\begin{eqnarray}
    \label{ESiSj}
    (k_{-1} + 2 k_{d}+ k_{+2})[E(S_{i}S_{j})^{e}] = 
    k_{+1}[E][(S_{i}S_{j})^{e}] + k_{a}([ES_{i}^{e}][S_{j}^{e}] + [ES_{j}^{e}][S_{i}^{e}]).
\end{eqnarray}
In order to get the speed of ligation, we have to rewrite $[E(S_{i}S_{j})^{e}]$ by $[E]$, $[S_{i}]$ and $[S_{j}]$.
To solve Eq.(\ref{ESiSj}), we need to obtain explicit forms of $[S_{i} S_{j}]$ and $[ES_{i}^{e}]$. 
First, according to Eqs. (\ref{FormESie})(\ref{ESiplusSj})(\ref{mismatchHyb}), $[ES_{i}^{e}]$,at steady state is described by,
\begin{eqnarray}
    \label{ESie}
    [ES_{i}^{e}] = \frac{
                            k_{+1} [E][S_{i}^{e}] 
                            + k_{d} \sum_{(i,l)\in\mathbb{P}}[ES_{i}^{e}S_{l}^{e}] 
                            + k_{d^{\prime}}\sum_{(i,l)\in\mathbb{I}}[ES_{i}^{e}S_{l}]
                        }
                        {
                            k_{-1} 
                            + k_{a}\sum_{(i,l)\in\mathbb{P}}[S_{l}^{e}] 
                            + k_{a^{\prime}}\sum_{(i,l)\in\mathbb{I}}[S_{l}]
                        }  .
\end{eqnarray}
For a dilute solution with $ \forall [S_{i}] \in \bm{S} \ll 1$ 
higher order terms such as $[S_{i}]^{3}$ are sufficiently small to assume in Eq.(\ref{ESiSj}) $ k_{a}\sum_{(i,l)\in\mathbb{P}}[S_{l}^{e}] 
+ k_{a^{\prime}}\sum_{(i,l)\in\mathbb{I}}[S_{l}] \ll k_{-1}$. One can find that the denominator in the right side of Eq. (\ref{ESie}) becomes constant $k_{-1}$.
The approximation of low concentration limit, local balance among  $[E]$, $[S_{i}^{e}]$, and $[ES_{i}^{e}]$ is given by
\begin{eqnarray}\label{eq14}
    [ES_{i}^{e}] = \frac{k_{+1}}{k_{-1}} [E][S_{i}^{e}] .
\end{eqnarray}
The same procedure also leads the local balance among $[(S_{i}S_{j})^{e}]$, $[S_{i}^{e}]$, and $[S_{j}^{e}]$ from Eq(\ref{normalHybDis}) as  
\begin{eqnarray}\label{eq15}
    k_{d}[(S_{i}S_{j})^{e}] = k_{a}[S_{i}^{e}][S_{j}^{e}]  .
\end{eqnarray}
By using Eqs. (\ref{eq14}) and (\ref{eq15}), Eq. (\ref{ESiSj}) is simplified as
\begin{eqnarray}
	\label{simpleESiSj}
    [E(S_{i}S_{j})^{e}] 
    = [E]
    \frac{
        k_{+1}k_{a}
        }
        {
        k_{-1} + 2k_{d} + k_{+2}
        }
         \Bigl(\frac{1}{k_{d}} + \frac{1}{k_{-1}}\Bigr)[S_{i}^{e}][S_{j}^{e}]  .
\end{eqnarray}

This expression is known as Lineweaver-Burk plot for enzymatic reaction in particular for DNA ligation.

\subsection*{The reaction pathway of DNA ligation in crowded solution}

Next, we give the explicit form of active ligase concentration $[E]$ as a function of total amount of ligase and substrate DNA concentrations in order to solve the speed of DNA ligation in crowded environment. Because the total amount of the enzyme is constant $[E_0]$, the fraction of the complexes made of ligase and DNA is given by 
\begin{eqnarray}
    \label{conservation}
    [E]_{0} 
    = [E] 
    + \sum_{(l,m)\in\mathbb{P}} 
        ( [E(S_{l}S_{m})^{e}] + [E \overline{S_{l}S_{m}}]) 
    + \sum_{(l,m)\in\mathbb{I}} 
        [ES_{l}S_{m}] 
    + \sum_{k}
        ( [ES_{k}^{e}] + [E\overline{S_{k}}] )
\end{eqnarray}

where $[E]$ is the freely available ligase in solution, $\sum_{(l,m)\in\mathbb{P}}$ is the summation over all possible complementary pairing of DNA, and $\sum_{(l,m)\in\mathbb{I}}$ is the summation over all partial complementary pairing of DNA

Given that the association/dissociation kinetics of ligase and DNA strands is faster than the chemical reaction, each terms for the ligase-DNA complexes in Eq.(\ref{conservation}) are rewritten as
\begin{eqnarray}\label{eq18}
    \sum_{n}
        ( [ES_{n}^{e}] + [E\overline{S_{n}}] ) 
    &=& \frac{k_{+1}}{k_{-1}} 
        [E] 
        \sum_{n}
            ( [S_{n}^{e}] + [\overline{S_{n}}] )    \nonumber \\
    &=& \frac{k_{+1}}{k_{-1}} 
        [E] 
        \sum_{n}
            (M(S_{n};l_{L})+1) [S_{n}]
\end{eqnarray}
from Eq. (\ref{nonHomologous})(\ref{eq14}).

Using Eq.(\ref{nonHomologous})(\ref{endLigation}), the kinetic production of complex $[ES_{l}S_{m}] ((l,m)\in\mathbb{I})$ in Eq.(\ref{eq18}), which is  complex of ligase and partial complementarily  hybridized DNA pairing, is described by,
\begin{eqnarray}
	\label{timeEvolESlSmInI}
    \frac{\mathrm{d}}{\mathrm{d} t} [ES_{l}S_{m}]
     &=& k_{a'}([ES_{l}][S_{m}] + [S_{l}][ES_{m}]) 
        + k_{+1}[E] (M(S_{l}S_{m};l_{L})+1) [S_{l}S_{m}] 
        - (k_{-1}+2k_{d^{\prime}})[ES_{l}S_{m}]             \nonumber \\
     &=& k_{+1}k_{a^{\prime}}[E] 
        \{
            k_{-1}^{-1}(M(S_{l};l_{L}) + M(S_{m};l_{L}) + 2) 
            + k_{d^{\prime}}^{-1}(M(S_{l}S_{m};l_{L})+1) 
        \} [S_{l}][S_{m}]                           \nonumber \\
     && - (k_{-1}+2k_{d^{\prime}})[ES_{l}S_{m}]  .
\end{eqnarray}
because we don’t have to consider homologous sequences of $S_{m}$ and $S_{l}$ $((l,m)\in\mathbb{I})$, length of $S_{m} = M(S_{m};l_{L})+1$.

Assuming steady-state of $ \frac{\mathrm{d}}{\mathrm{d} t} [ES_{l}S_{m}]
=0$, Eq. (\ref{timeEvolESlSmInI}) is simplified as,
\begin{eqnarray}
	\label{stableESlmInI}
    \sum_{(l,m)\in\mathbb{I}}
        [ES_{l}S_{m}]
     = &&[E] \frac{k_{+1}k_{a^{\prime}}}{k_{-1}+2k_{d^{\prime}}} \nonumber  \\
     \times 
        &&\sum_{(l,m)\in\mathbb{I}} 
        \{ 
            k_{-1}^{-1}(M(S_{l};l_{L}) + M(S_{m};l_{L}) + 2) + k_{d^{\prime}}^{-1}(M(S_{l}S_{m};l_{L})+1) 
        \}[S_{l}][S_{m}]
\end{eqnarray}
On the other hand, the complex $[E\overline{S_{i}S_{j}}]((i,j)\in\mathbb{P})$ is a complex of ligase $[E]$ and an non-homologous region of $S_{i}S_{j}$. From Eq.(\ref{nonHomologous})(\ref{normalHybDis}), its kinetics is represented by
\begin{eqnarray}
	\label{eqOfEbarSij}
    \frac{\mathrm{d}}{\mathrm{d} t} [E\overline{S_{i}S_{j}}]
     &=& k_{+1}[E] M(S_{i}S_{j};l_{L}) [\overline{S_{i}S_{j}}] 
         + k_{a}([E\overline{S_{i}}][S_{j}] + [S_{i}][E\overline{S_{j}}]) 
         - (k_{-1} + 2 k_{d}) [E\overline{S_{i}S_{j}}]          \nonumber \\
   &=& [E] k_{+1}k_{a} 
        \Bigl{\{} 
            \frac{1}{k_d} M(S_{i}S_{j};l_{L}) 
            + \frac{1}{k_{-1}} (M(S_{i};l_{L})+M(S_{i};l_{L}))  
        \Bigl{\}} [S_{i}][S_{j}] \nonumber \\
     && - (k_{-1} + 2 k_{d}) [E\overline{S_{i}S_{j}}]   .
\end{eqnarray}
By taking similar approximation if the system has reached steady state, that is $\frac{\mathrm{d}}{\mathrm{d} t} [E\overline{S_{i}S_{j}}]=0$, Eq. (\ref{eqOfEbarSij}) is also simplified as
\begin{eqnarray}
	\label{stableEbarSij}
    [E\overline{S_{i}S_{j}}] 
    = [E] \frac{k_{+1}k_{a}}{k_{-1} + 2 k_{d}} 
    \Bigl{\{} 
        \frac{1}{k_d} M(S_{i}S_{j};l_{L}) 
        + \frac{1}{k_{-1}} 
        \Bigl{(}
            M(S_{i};l_{L}) + M(S_{i};l_{L}) 
        \Bigl{)}  
    \Bigl{\}} [S_{i}][S_{j}]        .
\end{eqnarray}
By substituting $[E\overline{S_{i}S_{j}}]((i,j)\in\mathbb{P})$(Eq.(\ref{stableEbarSij})),  $[ES_{l}S_{m}]((l,m)\in\mathbb{I})$(Eq.(\ref{stableESlmInI})) and  $[ES_{n}^{e}]+[E\overline{S_{n}}]$(Eq.(\ref{eq18})) into Eq. (\ref{conservation}), the freely available ligase concentration is
\begin{eqnarray}
	\label{FinalFreeE}
    [E] 
    &=& ([E]_{0} - \sum_{(l,m)\in\mathbb{P}} [E(S_{l}S_{m})^{e}]) \nonumber \\
    & &\times 
    \Bigl{\{} 
        1 +  \frac{k_{+1}k_{a^{\prime}}}{k_{-1}+2k_{d^{\prime}}} 
        \sum_{(l,m)\in\mathbb{I}} 
            \Bigl{\{} 
                k_{-1}^{-1}(M(S_{l};l_{L}) + M(S_{m};l_{L}) + 2) 
                + k_{d^{\prime}}^{-1}(M(S_{l}S_{m};l_{L})+1) 
            \Bigl{\}}[S_{l}][S_{m}] \nonumber \\
        && + \frac{k_{+1}k_{a}}{k_{-1} + 2 k_{d}} 
        \sum_{(i,j)\in\mathbb{P}} 
            \Bigl{\{} 
                k_{-1}^{-1} 
                    \Bigl{(}
                        M(S_{i};l_{L}) + M(S_{i};l_{L}) 
                    \Bigl{)}
                + k_{d}^{-1} M(S_{i}S_{j};l_{L}) 
            \Bigl{\}} [S_{i}][S_{j}] \nonumber \\
        && + \frac{k_{+1}}{k_{-1}}
            \sum_{n}
                \Bigl{(} 
                    (M(S_{n};l_{L})+1) [S_{n}] 
                \Bigl{)}  
    \Bigl{\}}^{-1}                   \nonumber \\
    &=& ([E]_{0} - \sum_{(l,m)\in\mathbb{P}} [E(S_{l}S_{m})^{e}]) \Theta(\textbf{S})^{-1}
\end{eqnarray}
where
\begin{eqnarray}
    \Theta(\textbf{S}) &\equiv& 
    1 
    + \frac{k_{+1}k_{a^{\prime}}}{k_{-1}+2k_{d^{\prime}}} 
    \sum_{(l,m)\in\mathbb{I}} 
        \Bigl{\{} 
            k_{-1}^{-1}(M(S_{l};l_{L}) + M(S_{m};l_{L}) + 2) 
            + k_{d^{\prime}}^{-1}(M(S_{l}S_{m};l_{L})+1) 
        \Bigl{\}}[S_{l}][S_{m}]         \nonumber \\
    && + \frac{k_{+1}k_{a}}{k_{-1} + 2 k_{d}} 
        \sum_{(i,j)\in\mathbb{P}} 
            \Bigl{\{} 
                k_{-1}^{-1} \Bigl{(}M(S_{i};l_{L}) + M(S_{i};l_{L}) \Bigl{)}
                + k_{d}^{-1} M(S_{i}S_{j};l_{L}) 
            \Bigl{\}} [S_{i}][S_{j}]    \nonumber \\
    && + \frac{k_{+1}}{k_{-1}}
        \sum_{n} \Bigl{(} 
            (M(S_{n};l_{L})+1) [S_{n}] 
        \Bigl{)}        .
\end{eqnarray}
By inserting Eq.(\ref{FinalFreeE}) into Eq.(\ref{simpleESiSj}), we obtain
\begin{eqnarray}
    \label{answer1}
    [E(S_{i}S_{j})^{e}] 
    = \Psi^{-1} [S_{i}^{e}][S_{j}^{e}]
      ([E]_{0} - \sum_{(l,m)\in\mathbb{P}} [E(S_{l}S_{m})^{e}]) \Theta(\textbf{S})^{-1}
\end{eqnarray}
where
\begin{eqnarray}
    \Psi^{-1} \equiv \frac{k_{+1}k_{a}}{k_{-1} + 2k_{d} + k_{+2}} (\frac{1}{k_{d}} + \frac{1}{k_{-1}})  .
\end{eqnarray}
The summation of eq.(\ref{answer1}) over all possible complementary pairing of DNA is given by,
\begin{eqnarray}
    \sum_{(i,j)\in\mathbb{P}} [E(S_{i}S_{j})^{e}] 
    = \Psi^{-1} \Theta(\textbf{S})^{-1}
    ([E]_{0} - \sum_{(l,m)\in\mathbb{P}} [E(S_{l}S_{m})^{e}])
    \sum_{(i,j)\in\mathbb{P}}[S_{i}^{e}][S_{j}^{e}] .
\end{eqnarray}
We solve this equation and obtain the following expression
\begin{eqnarray}
    \label{sum}
    \sum_{(l,m)\in\mathbb{P}} [E(S_{l}S_{m})^{e}]
    = \frac{
        [E]_{0} \sum_{(i,j)\in\mathbb{P}}[S_{i}^{e}][S_{j}^{e}]
    }{
        \Psi \Theta(\textbf{S}) 
        + \sum_{(i,j)\in\mathbb{P}}[S_{i}^{e}][S_{j}^{e}]
    }
\end{eqnarray}
Substituting Eq. (\ref{defOfSe})(\ref{sum}) into Eq. (\ref{answer1}) gives
\begin{eqnarray}
    [E(S_{i}S_{j})^{e}] 
    = \frac{
            [E]_{0}[S_{i}][S_{j}]
        }{
            \Psi \Theta(\textbf{S})
            + \sum_{(l,m)\in\mathbb{P}}[S_{l}][S_{m}]
        }      .
\end{eqnarray}
Finally, the speed of producing $S_{k}$ from $S_{i}$ and $S_{j}$ is given by
\begin{eqnarray}
    v_{ij}(\textbf{S}) 
    \equiv \frac{\mathrm{d}}{\mathrm{d}t}[S_{k}]
    = k_{+2}[E(S_{i}S_{j})^{e}] 
    = \frac{
        k_{+2}[E]_{0}[S_{i}][S_{j}]
        }{
           \Psi \Theta(\textbf{S})
            + \sum_{(l,m)\in\mathbb{P}}[S_{l}][S_{m}]
        }      
\end{eqnarray}
where $v_{ij}(\textbf{S})$ is the speed of ligation of $S_{i}+S_{j} \to S_{k}$.

\section{Dependence of kinetic parameters on the concentration of crowding agent}
In this section, we present theoretical description of the dependence on the concentration of PEG, $c_{peg}$.

First, we consider the crowding induced change of the affinity between DNA fragments. The thermodynamic stability of DNA duplex is important to hybridize sticky ends of DNA fragments. It has been known that crowding agents such as PEG decreases a melting temperature, $T_{m}$, of short DNA fragment \cite{Nakano}. The fraction of duplex DNA is decided by the thermodynamic stability that decreases at $T_{m}$. Therefore the fraction of duplex DNA, $r(T, c_{peg})$, is given by
Crowding effects enhance the binding affinity of DNA fragments by entropic force. According to previous study by Minton, et al\cite{Minton}, the activity of enzyme is given by the following equation as a function of the polymer concentration,
\begin{eqnarray}
    r(T,c_{ peg }) &=& \frac{1}{1+\exp((T-(T_{m} - c_{ peg }/3))/10)},
\end{eqnarray}
where $T$ is temperature in bulk solution and $T_{m}$ is melting temperature of complementary 4bp of sticky end of DNA\cite{Marks} and the dissociation constants $K_{d} and  K_{d}^{\prime}$ in the hybridization of sticky ends are proportional to $[S_{i}][S_{j}] /[S_{i}S_{j}] \sim 1/r(T, c_{ peg })$. By using this relation, those kinetic constants are reduced to
\begin{eqnarray}
    K_{d} &=&   10^4 \times \frac{1+\exp((T-(T_{m} - c_{peg}/3))/10)}{1+\exp((T-T_{m})/10)} \exp(-(c_{peg})^2) \\
    K_{d^{\prime}} &=&  10^6 \times \frac{1+\exp((T-(T_{m}^{\prime} - c_{peg}/3))/10)}{1+\exp((T-T_{m}^{\prime})/10)} \exp(-(c_{peg})^2)
\end{eqnarray}
with $T_{m}^{\prime}$ melting temperature of non-complementary 4bp of sticky end. In this study, $T$=\SI{25}{\celsius}, $T_{m}$=\SI{12}{\celsius} and $T_{m}^{\prime}$=\SI{4}{\celsius}. 

Both DNA hybridization and DNA-ligase complex formation are given by number density of chemical species, and the dissociation is regulated by interaction forces.
Therefore, the association constant $k_a$ and $k_{a^{\prime}}$, and the dissociation constants $k_d$ and $k_{d^{\prime}}$ are given by
\begin{eqnarray}
    k_{a} &=& 10^0 \\
    k_{d} &=& K_{d}k_{a} \\
    k_{a^{\prime}} &=& 10^0 \\
    k_{d^{\prime}} &=& K_{d^{\prime}}k_{a^{\prime}}     .
\end{eqnarray}

$ $

Second, the affinity constant between DNA and ligase is considered Macromolecular crowding suppresses the dissociation of enzyme from bound DNA. The binding constant $k_{+1}$ and the dissociation constant $k_{-1}$ are given by
\begin{eqnarray}
    k_{+1} &=& 10^0 \\
    k_{-1} &=& k_{+1}10^2 \exp(-(c_{peg}/10)^2)  .
\end{eqnarray}

$ $

Finally, we give explicit form of the production rate of ligated DNA $k_{+2}$ .It has been known that Taq DNA ligase catalyze dehydration reaction with coenzyme. In contrast, crowding agent such as PEG attracts water molecules due to its hydrophilic nature, and the ligation reaction slows down because the fraction of freely available water could be reduced in higher PEG concentration. We assume that the speed of ligation decreases exponentially with the concentration of PEG\cite{homchaudhuri}, the kinetic constant $k_{+2}$ for ligation is defined by
\begin{eqnarray}
    k_{+2} &=& 10^{3-\frac{c_{peg}}{2}}. 
\end{eqnarray}

\end{document}